\documentclass[a4paper,11pt]{article}

\usepackage{jheppub}
\usepackage{units}
\usepackage{xspace}
\usepackage{mathrsfs}
\usepackage{units}
\usepackage{amssymb}
\usepackage{placeins}
\usepackage{subfigure}
\usepackage{chngcntr}
\counterwithout{equation}{section}

\newcommand{\Alpgen}{{\tt ALPGEN}\xspace}
\newcommand{\Jimmy}{{\tt JIMMY}\xspace}
\newcommand{\Mcatnlo}{{\tt MC@NLO}\xspace}
\newcommand{\Sherpa}{{\tt SHERPA}\xspace}
\newcommand{\Herwig}{{\tt HERWIG}\xspace}
\newcommand{\Herwigpp}{{\tt Herwig++}\xspace}

\newcommand{\Prospino}{{\tt PROSPINO}\xspace}
\newcommand{\Madgraph}{{\tt MadGraph}\xspace}

\newcommand{\CTEQSixL}{{\tt CTEQ6L1}\xspace}
\newcommand{\CTEQSix}{{\tt CTEQ6.6}\xspace}
\newcommand{\CTTen}{{\tt CT10}\xspace}
\newcommand{\PYTHIA}{{\tt PYTHIA}\xspace}
\newcommand{\GEANT}{{\tt GEANT4}\xspace}
\newcommand{\MSTW}{{\tt MSTW}\xspace}

\newcommand{\MCFM}{{\tt MCFM}\xspace}
\newcommand{\fewz}{{\tt FEWZ}\xspace}

\newcommand{\POWHEG}{{\tt POWHEG}\xspace}
\newcommand{\antibar}[1]{\ensuremath{#1\bar{#1}}}
\newcommand{\ttbar}{\antibar{t}\xspace}

\newcommand{\stau}{\ensuremath{\tilde{\tau}_1}\xspace}

\newcommand{\ninoone}{\ensuremath{\tilde{\chi}_1^0}\xspace}

\newcommand{\chinoone}{\ensuremath{\tilde{\chi}_1^\pm}\xspace}

\newcommand{\met}{\ensuremath{E_{\mathrm{T}}^{\mathrm{miss}}}\xspace}

\def\ninoi{\ensuremath{\mathchoice
      {\displaystyle\raise.4ex\hbox{$\displaystyle\tilde\chi^0_j$}}%
         {\textstyle\raise.4ex\hbox{$\textstyle\tilde\chi^0_j$}}%
       {\scriptstyle\raise.3ex\hbox{$\scriptstyle\tilde\chi^0_j$}}%
 {\scriptscriptstyle\raise.3ex\hbox{$\scriptscriptstyle\tilde\chi^0_j$}}}}

\def\chinojpm{\ensuremath{\mathchoice%
      {\displaystyle\raise.4ex\hbox{$\displaystyle\tilde\chi^\pm_i$}}%
         {\textstyle\raise.4ex\hbox{$\textstyle\tilde\chi^\pm_i$}}%
       {\scriptstyle\raise.3ex\hbox{$\scriptstyle\tilde\chi^\pm_i$}}%
{\scriptscriptstyle\raise.3ex\hbox{$\scriptscriptstyle\tilde\chi^\pm_1$}}}}

\usepackage{preprintcover}  
\PreprintCoverPaperTitle{Search for $R$-parity-violating supersymmetry in events with four or more leptons in $\sqrt{s}\,=\,$7$\,$TeV $pp$ collisions with the ATLAS detector}  
\PreprintIdNumber{CERN-PH-EP-2012-276}  
\PreprintCoverAbstract{A search for new phenomena  in final states with four or more leptons (electrons or muons) is presented.
The analysis is based on 4.7$\,$fb$^{-1}$ of $\sqrt{s}\,=\,$7$\,$TeV proton-proton collisions delivered by the Large Hadron Collider and recorded with the ATLAS detector.
Observations are consistent with Standard Model expectations in two signal regions: one that requires moderate values of missing transverse momentum and another that requires large effective mass. 
The results are interpreted in a simplified model of $R$-parity-violating supersymmetry in which a 95\% CL exclusion region is set for charged wino masses up to 540$\,$GeV. In an $R$-parity-violating MSUGRA/CMSSM model, values of $m_{1/2}$ up to 820$\,$GeV are excluded for $10<\tan\beta<40$.}
\PreprintJournalName{JHEP}  

\begin{document}

 \title{\boldmath Search for $R$-parity-violating supersymmetry in events with four or more leptons in $\sqrt{s}\,=\,$7$\,$TeV $pp$ collisions with the ATLAS detector}

\author{The ATLAS Collaboration}

\abstract{ 
A search for new phenomena  in final states with four or more leptons (electrons or muons) is presented.
The analysis is based on 4.7$\,$fb$^{-1}$ of $\sqrt{s}\,=\,$7$\,$TeV proton-proton collisions delivered by the Large Hadron Collider and recorded with the ATLAS detector.
Observations are consistent with Standard Model expectations in two signal regions: one that requires moderate values of missing transverse momentum and another that requires large effective mass. 
The results are interpreted in a simplified model of $R$-parity-violating supersymmetry in which a 95\% CL exclusion region is set for charged wino masses up to 540$\,$GeV. In an $R$-parity-violating MSUGRA/CMSSM model, values of $m_{1/2}$ up to 820$\,$GeV are excluded for $10<\tan\beta<40$. 
}

\maketitle
\flushbottom

\section{Introduction}

Events with four or more leptons are rarely produced in Standard Model (SM) processes while being predicted by a variety of theories for physics beyond the SM. These include 
supersymmetry (SUSY)~\cite{Miyazawa:1966,Ramond:1971gb,Golfand:1971iw,Neveu:1971rx,Neveu:1971iv,Gervais:1971ji,Volkov:1973ix,Wess:1973kz,Wess:1974tw},  technicolour~\cite{technicolor}, and models with extra bosons~\cite{extraBoson} or heavy neutrinos~\cite{heavyNeutrino}.  This paper presents a search with the ATLAS detector for anomalous production of events with four or more leptons.  ``Leptons'' refers to electrons or muons, including those from $\tau$ decays, but does not include $\tau$ leptons that decay hadronically.  
The analysis is based on 4.7$\,$fb$^{-1}$ of proton-proton collisions delivered by the Large Hadron Collider (LHC) at a centre-of-mass energy $\sqrt{s}\,$$=\,$7$\,$TeV between March and October 2011. The results are interpreted in the context of $R$-parity-violating (RPV) SUSY. 
Similar searches have been conducted at the Tevatron Collider~\cite{D02006, CDF2007} and by CMS~\cite{CMS4L}.

\section{$R$-parity-violating supersymmetry}
SUSY postulates the existence of SUSY particles, or ``sparticles'', each with spin ($S$) differing by one-half unit from that of its SM partner.
Gauge-invariant and renormalisable interactions introduced in SUSY models can violate the conservation of baryon ($B$) and lepton ($L$) number and lead to a proton lifetime shorter than current experimental limits~\cite{protonDecay}.
This is usually solved by assuming that $R$-parity, defined by $P_R=(-1)^{2S+3B+L}$, is conserved ~\cite{Fayet:1976et,Fayet:1977yc,Farrar:1978xj,Fayet:1979sa,Dimopoulos:1981zb},  
which makes the lightest supersymmetric particle (LSP) stable.
In $P_R$-conserving models where the LSP is neutral and weakly interacting, sparticle production is characterised by large missing transverse momentum (\met) due to LSPs escaping detection. Many SUSY searches at hadron colliders rely on this large $\met$ signature.

Alternatively, proton decay can be prevented by imposing other symmetries~\cite{herbiMSSMsymm} that require the conservation of either lepton or baryon number, while allowing $R$-parity violation.
Such models can accommodate non-zero neutrino masses and neutrino-mixing angles consistent with the observation of neutrino oscillations~\cite{neutrinomass_benall}. 
If $R$-parity is violated, the LSP decays into SM particles and the signature of large \met may be lost.
Trilinear lepton-number-violating RPV interactions can generate both charged leptons and neutrinos during the LSP decay, and therefore lead to a characteristic signature with high lepton multiplicity and moderate values of \met compared to $R$-parity-conserving models.

\section{$R$-parity-violating models}

The results of this analysis are interpreted in two Minimal Supersymmetric Standard Model (MSSM) scenarios with an RPV superpotential term given by $W_{RPV} = \lambda_{ijk}L_iL_j\bar{E_k}$. The $i, j, k$ indices of the $\lambda_{ijk}$ Yukawa couplings refer to the lepton generations. The lepton  SU(2) doublet superfields are denoted by $L_i$, while the corresponding singlet superfields are given by $E_k$.
Single coupling dominance is assumed with $\lambda_{121}$ as the only non-zero coupling.
The $\lambda_{121}$ coupling is chosen as a representative model with multiple electrons and muons in the final state. Comparable signal yields are expected with a choice of $\lambda_{122}$ single coupling dominance.

\begin{figure}[tb]
\centering
\subfigure[\label{fig:neutralinoDecay1}]{\includegraphics[height=2.5cm]{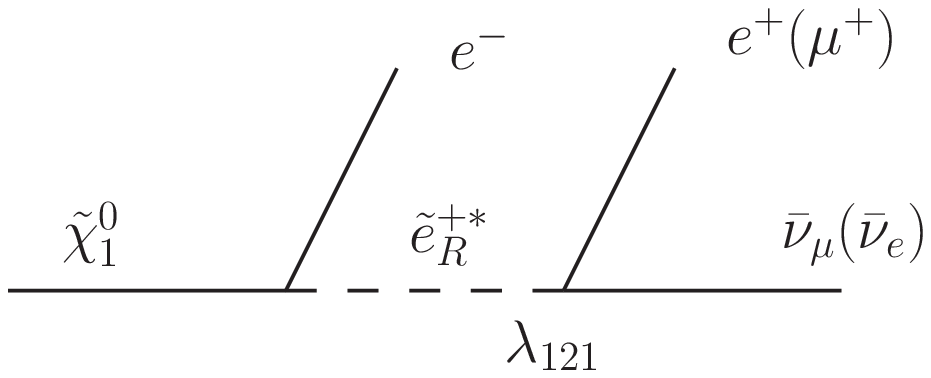}}
\hspace{1.5cm}
\subfigure[\label{fig:neutralinoDecay2}]{\includegraphics[height=2.5cm]{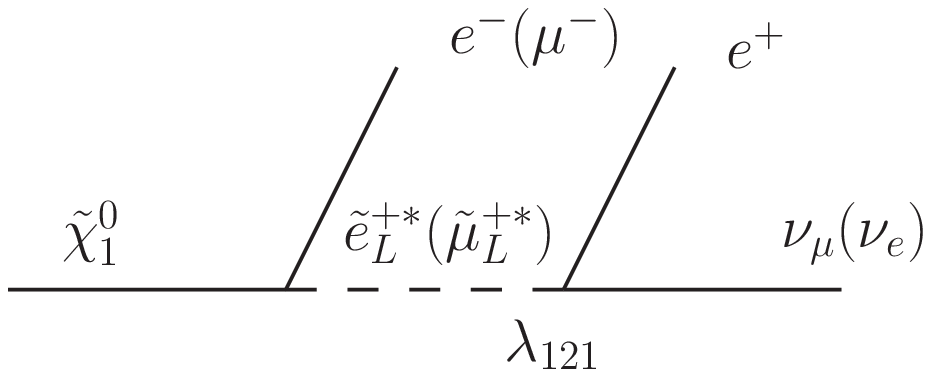}}
\subfigure[\label{fig:neutralinoDecay3}]{\includegraphics[height=2.5cm]{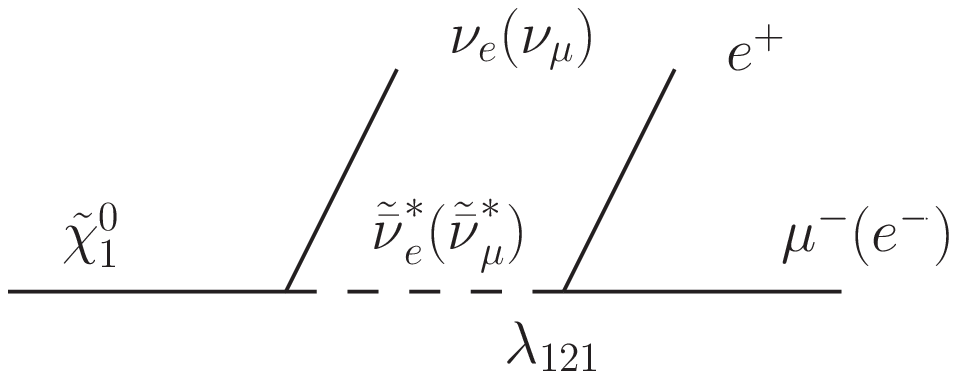}}
\hspace{0.8cm}
\subfigure[\label{fig:stauDecay}]{\includegraphics[height=2.5cm]{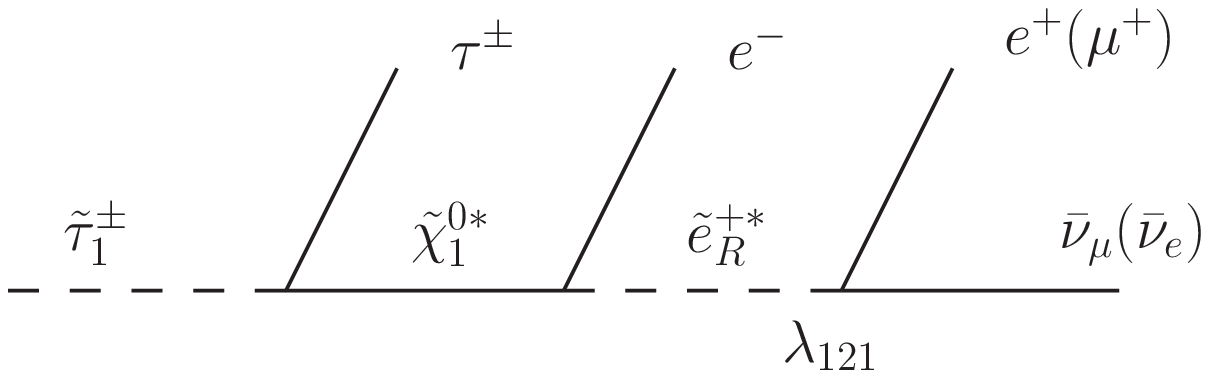}}
\caption{Illustration of the $\ninoone$ decay in the chosen simplified model (a--c), and the four-body $\stau$ decay in the MSUGRA/CMSSM model (d). The charge conjugate decay of the $\ninoone$ decay is implied.\label{fig:staufeynman} }
\end{figure}

The first scenario is a simplified model~\cite{simplModels1} where the lightest chargino and neutralino are the only sparticles with masses below the TeV scale.
Charginos ($\chinojpm$, $i=1,2$) and neutralinos ($\ninoi$, $j=1,2,3,4$) are the mass eigenstates formed from the linear superposition of the SUSY partners of the Higgs and electroweak gauge bosons.
These are the Higgsinos, winos and bino.
Wino-like charginos are pair-produced and each decays into a $W$ boson and a bino-like $\ninoone$, which is the LSP.
The LSP then undergoes the three-body decay $\ninoone\rightarrow e \mu \nu_{e}$ or $\ninoone\rightarrow e e \nu_{\mu}$ through a virtual slepton or sneutrino, with a branching fraction of 50\% each, as shown in figure~\ref{fig:staufeynman}(a--c).
The width of the $\ninoone$ is fixed at a value of 100$\,$MeV resulting in prompt decays.

The second scenario is taken from ref.~\cite{Desch:2011} and is the ($m_{1/2}$, $\tan\beta$) slice of the minimal SUper GRAvity/Constrained MSSM (MSUGRA/CMSSM) containing the BC1 point of ref.~\cite{bintosgamma}. The unification parameters $m_0$ and $A_0$ are zero and $\mu$ is positive\footnote{The parameter $m_0$ is the universal scalar mass, $m_{1/2}$ is the universal gaugino mass, $\tan\beta$ is the ratio of the two Higgs vacuum expectation values in the MSSM, $A_0$ is the trilinear coupling and $\mu$ is the Higgs mixing parameter, all defined at the unification scale.}. The RPV coupling $\lambda_{121}$ is set to 0.032 at the unification scale~\cite{Kao:2009fg}.
Both strong and weak processes contribute to SUSY pair production, where weak processes are dominant above ${m_{1/2}\,\sim\,600\,}$GeV.
The lighter stau, $\stau$, is the LSP over most of the parameter-space, and decays with equal probability through the four-body decays 
$\stau\rightarrow \tau e \mu \nu_{e}$ or $\stau\rightarrow \tau e e \nu_{\mu}$, via a virtual neutralino and sneutrino or slepton, as shown in figure~\ref{fig:stauDecay}. 
While the Higgs mass values in this model are lower than those of the recently observed Higgs-like resonance~\cite{HiggsATLAS,HiggsCMS}, the MSUGRA/CMSSM scenario considered nevertheless remains an instructive benchmark model. 

\section{Detector description}

ATLAS~\cite{atlas-det} is a multipurpose particle detector with forward-backward symmetric cylindrical geometry.
It includes an inner tracker (ID) immersed in a 2$\,$T axial magnetic field providing precision tracking of charged particles for pseudorapidities\footnote{ATLAS uses a right-handed coordinate
  system with its origin at the nominal interaction point (IP) in the
  centre of the detector and the $z$-axis along the beam pipe. The
  $x$-axis points from the IP to the centre of the LHC ring, and the
  $y$-axis points upward. Cylindrical coordinates $(r,\phi)$ are used
  in the transverse plane, $\phi$ being the azimuthal angle around the
  beam pipe. The pseudorapidity is defined in terms of the polar angle
  $\theta$ as $\eta=-\ln\tan(\theta/2)$.} $|\eta|\,$$<\,$2.5. 
Sampling calorimeter systems with either liquid argon or scintillator tiles as the active media provide energy measurements over the range $|\eta|\,$$<\,$4.9.
The muon detectors are positioned outside the calorimeters and are contained in an air-core toroidal magnetic field produced by superconducting magnets with field integrals varying from 1$\,$T$\cdot$m to 8$\,$T$\cdot$m.
They provide trigger and high-precision tracking capabilities for $|\eta|\,$$<\,$2.4 and $|\eta|\,$$<\,$2.7, respectively.

\section{Monte Carlo simulation}

Several Monte Carlo (MC) generators are used to simulate SM processes and new physics signals relevant for this analysis.
\Sherpa~\cite{SherpaMC} is used to simulate the diboson processes $WW$, $WZ$ and $ZZ$, where $Z$ also includes  virtual photons.
These diboson samples correspond to all SM diagrams leading to the $\ell\nu\ell'\nu'$, $\ell\ell\ell'\nu'$, and $\ell\ell\ell'\ell'$ final states, where $\ell,\ell'=e,\mu,\tau$ and $\nu,\nu'=\nu_e,\nu_{\mu},\nu_{\tau}$. Interference between the diagrams is taken into account. 
\Madgraph~\cite{madgraph} is used for the $t\bar{t}W$,  $t\bar{t}WW$, $t\bar{t}Z$, $W\gamma$ and $Z\gamma$ processes.
\Mcatnlo~\cite{mcatnlo} is chosen for the simulation of single and pair production of top quarks, and \Alpgen~\cite{alpgen} is used to simulate $W$+jets and $Z$+jets processes.
Expected diboson yields are normalised using next-to-leading-order (NLO) QCD predictions obtained with \MCFM~\cite{mcfm1,mcfm2}.
The top-quark pair-production contribution is normalised to approximate next-to-next-to-leading-order calculations (NNLO)~\cite{hathor} and the $t\bar{t}W$, $t\bar{t}WW$, $t\bar{t}Z$ contributions are normalised to NLO predictions~\cite{ttZ, ttW}.
The $W\gamma$ and $Z\gamma$  yields are normalised to be consistent with the ATLAS cross-section measurements~\cite{wgamma}.
The QCD NNLO \fewz~\cite{fewz1,fewz2} cross-sections are used for normalisation of the inclusive $W$+jets and $Z$+jets processes.

The choice of the parton distribution functions (PDFs) depends on the generator.
The \CTEQSixL~\cite{CTEQ6L1} PDFs are used with \Madgraph\ and \Alpgen, and the \CTTen~\cite{CT10pdf} PDFs with \Mcatnlo and \Sherpa.

The simplified model samples are produced  with \Herwigpp~\cite{herwigplusplus} and the MSUGRA/ CMSSM BC1-like samples are produced with \Herwig~\cite{herwig}.
The yields of the SUSY samples are normalised to the NLO cross-sections obtained from \Prospino~\cite{Beenakker:1996ch} for weak processes, and to next-to-leading-logarithmic accuracy (NLL) for strong processes.

Fragmentation and hadronisation for the \Alpgen and \Mcatnlo  samples are performed with \Herwig, while for \Madgraph, \PYTHIA~\cite{pythia} is used, and for \Sherpa these are performed internally.
\Jimmy~\cite{jimmy} is interfaced to \Herwig for simulation of the underlying event.  
For all MC samples, the propagation of particles through the ATLAS detector is modelled using \GEANT~\cite{Agostinelli:2002hh,simulation}. 
The effect of multiple proton-proton collisions from the same or different bunch crossings is incorporated into the simulation by overlaying additional minimum-bias events generated by \PYTHIA onto hard-scatter events. 
Simulated events are weighted to match the distribution of the number of interactions per bunch crossing observed in data.
Simulated data are reconstructed in the same manner as the data.

\section{Event reconstruction and preselection \label{sec:presel}}

The data sample was collected with an inclusive selection of single-lepton and double-lepton triggers.  
For single-lepton triggers, at least one reconstructed muon (electron) is required to have transverse momentum $p_{\rm T}^{\mu}$ (transverse energy $E_{\rm T}^e$) above 20$\,$GeV (25$\,$GeV). 
For dilepton triggers, at least two reconstructed leptons are required to have triggered the event, with transverse energy or momentum above threshold. 
The two muons are each required to have $p_{\rm T}^{\mu}\,$$>\,$12$\,$GeV for dimuon triggers, and the two electrons to have $E_{\rm T}^{e}\,$$>\,$17$\,$GeV for dielectron triggers, while the thresholds for electron-muon triggers are $E_{\rm T}^e\,$$>\,$15$\,$GeV and  $p_{\rm T}^{\mu}\,$$>\,$10$\,$GeV.
These thresholds are chosen such that the overall trigger efficiency is high, typically in excess of 90\%, and independent of the transverse momentum of the triggerable objects within uncertainties.
 
Events recorded during normal running conditions are analysed if the primary vertex has five or more tracks associated to it.
The primary vertex of an event is identified as the vertex with the highest $\Sigma p_{\rm T}^2$ of associated tracks.

Electrons must satisfy ``medium'' identification criteria~\cite{electronPerf} and fulfil $|\eta|\,$$<\,$2.47 and $E_{\rm T}\,$$>\,$10$\,$GeV, where $E_{\rm T}$ and $|\eta|$ are determined from the calibrated clustered energy deposits in the electromagnetic calorimeter and the matched ID track, respectively. 
Muons are reconstructed by combining tracks in the ID and tracks in the muon spectrometer~\cite{muon}.
Reconstructed muons are considered as candidates if they have transverse momentum $p_{\rm T}\,$$>\,$10$\,$GeV and $|\eta|\,$$<\,$2.4.

Jets are reconstructed with the  anti-$k_t$ algorithm~\cite{Cacciari:2008gp} with a radius parameter of ${R=0.4}$ using clustered energy deposits calibrated at the electromagnetic scale\footnote{The electromagnetic scale is the basic calorimeter signal scale for the ATLAS calorimeters. It has been established using test-beam measurements for electrons and muons to give the correct response for the energy deposited in electromagnetic showers, although it does not correct for the lower response of the calorimeter to hadrons.}.
The jet energy is corrected to account for the non-compensating nature of the calorimeter using correction factors parameterised as a function of the jet $E_{\rm T}$ and $\eta$~\cite{fakeCleanUp}.
The correction factors were obtained from simulation and have been refined and validated using data.
Jets considered in this analysis have $E_{\rm T}\,$$>\,$20$\,$GeV and $|\eta|\,$$<\,$2.5. 
The $p_{\rm T}$-weighted fraction of the tracks in the jet that are associated with the primary vertex is required to be larger than 0.75. 

Events containing jets failing the quality criteria described in ref.~\cite{fakeCleanUp} are rejected to suppress both SM and beam-induced background. 
Jets are identified as containing $b$-hadron decays, and thus called ``$b$-tagged'', using a multivariate technique based on quantities such as the impact parameters of the tracks associated to a reconstructed secondary vertex.  
The chosen working point of the $b$-tagging algorithm~\cite{btag} correctly identifies $b$-quark jets in simulated top-quark decays with an efficiency of 60\% and misidentifies the jets initiated by light-flavour quarks or gluons with a rate of $<\,$1\%, for jets with $E_{\rm T}\,$$>\,$20$\,$GeV and $|\eta|\,$$<\,$2.5.

The missing transverse momentum, $\met$, is the magnitude of the vector sum of the transverse momentum or transverse energy of all $p_{\rm T}\,$$>\,$10$\,$GeV muons, $E_{\rm T}\,$$>\,$20$\,$GeV electrons, $E_{\rm T}\,$$>\,$20$\,$GeV jets, and calibrated calorimeter energy clusters with $|\eta|\,$$<\,$4.9 not associated to these objects~\cite{metPerf}.

In this analysis, ``tagged'' leptons are leptons  separated from each other and from candidate jets as described below. 
If two candidate electrons are reconstructed with $\Delta R \equiv \sqrt{(\Delta\phi)^2+(\Delta\eta)^2}\,$$<\,$0.1, the lower energy one is discarded.  Candidate jets within $\Delta R\,$$=\,$0.2 of an electron candidate are rejected. To suppress leptons originating from  semi-leptonic decays of $c$- and $b$-quarks, all lepton candidates within $\Delta R\,$$=\,$0.4 of any remaining jet candidates are removed. 
Muons undergoing bremsstrahlung can  be reconstructed with an overlapping electron candidate. To reject these, tagged electrons and muons separated from jets and reconstructed within $\Delta R\,$$=\,$0.1 of each other are both discarded. 
Events containing one or more tagged muons that have transverse impact parameter with respect to the primary vertex $|d_0|\,$$>\,$0.2$\,$mm or longitudinal impact parameter with respect to the primary vertex $|z_0|\,$$>\,$1$\,$mm are rejected to suppress cosmic muon background.
  
``Signal'' leptons are tagged leptons for which the scalar sum of the transverse momenta of tracks within a cone of $\Delta R = 0.2$ around the lepton candidate, and excluding the lepton candidate track itself, is less than 10\% of the lepton $E_{\rm T}$ for electrons and less than 1.8$\,$GeV for muons. 
Tracks selected for the electron and muon isolation requirement defined above are those which have $p_{\rm T}\,$$>\,$1$\,$GeV and are associated to the primary vertex of the event. Signal electrons must also pass ``tight'' identification criteria~\cite{electronPerf}.  

\section{Signal region selection \label{sec:SRsel}}

Selected events must contain four or more signal leptons.
The invariant mass of any same-flavour opposite-sign (SFOS) lepton pair, $m_{\rm SFOS}$, must be above 20$\,$GeV, otherwise the lepton pair is discarded to suppress background from low-mass resonances. Events which contain a SFOS lepton pair inside the [81.2, 101.2]$\,$GeV interval are vetoed to reject $Z$-boson candidates ($Z$-veto).

Two signal regions are then defined: a signal region with $\met\,$$>\,$50$\,$GeV (SR1) and one with effective mass $m_{\rm eff}\,$$>\,$300$\,$GeV (SR2).
The effective mass is defined by the scalar sum shown in (eq. \ref{eq:meff}), where $p^\mu_{\rm T}$ ($E^e_{\rm T}$) is the transverse momentum of the signal muons (electrons) and  $E^j_{\rm T}$ is the transverse energy of jets with $E_{\rm T}\,$$>\,$40$\,$GeV: 

\begin{equation}
m_{\rm eff} = \met + \sum_{\mu} p^\mu_{\rm T} + \sum_{e} E^e_{\rm T} + \sum_{j} E^j_{\rm T}.
\label{eq:meff}
\end{equation}

SR1 is optimised for regions of the parameter space with values of $m_{1/2}$ below $\sim$700$\,$GeV or $m_{\chinoone}$ below $\sim$300$\,$GeV. SR2 targets the production of heavier sparticles.  In general, SR1 is sensitive to models with $\met$ originating from neutrinos and SR2 to scenarios with a large multiplicity of high-$p_{\rm T}$ objects.

Three further regions are defined to validate the expected background against data. The validation regions are described in section~\ref{sec:Bvalidation}.
Table~\ref{tab:SRdef} summarises the signal and validation region definitions.

\begin{table}[t]
\begin{center}
 \caption{The selection requirements for the two signal (SR) and three validation regions (VR).
 The $Z$-candidate veto (requirement) rejects (selects) events that have a same-flavour opposite-sign lepton pair with mass inside the [81.2, 101.2]$\,$GeV interval. \label{tab:SRdef}}
 \small{
 \begin{tabular}{c c c c c c}
  \hline\hline
  Selection		& SR1 & SR2 & VR1 & VR2 & VR3 \\
\hline
Number of leptons  & $\geq\,$4 & $\geq\,$4 & 3 & $\geq\,$4 & $\geq\,$4 \\ 
Z-candidate & veto & veto & veto &  requirement & veto \\
$E_{\rm T}^{\rm miss}$/GeV & $>\,$50 & -- & $>\,$50 & -- &  $<\,$50 \\ 
$m_{\rm eff}$/GeV  & -- & $>\,$300 & -- & -- &  $<\,$300 \\
\hline
\hline
 \end{tabular}
 }
 \end{center}
\end{table}

\section{Standard Model background estimation}

Several SM processes,   which are classified into irreducible and reducible components below, contribute to the background in the signal regions. 
The dominant sources are $ZZ$, $WZ$,  and $t\bar{t}$ production in both SR1 and SR2. 
 
\subsection{Irreducible background processes}

A background process is considered ``irreducible'' if it leads to events with four real, isolated leptons, referred to as ``real'' leptons below.
These include $ZZ$, $t\bar{t}Z$ and $t\bar{t}WW$ production, where a gauge boson may be produced off-mass-shell.
These  contributions are determined using the corresponding MC samples, for which lepton and jet selection efficiencies~\cite{JESuncert1,JERuncert1,LeptonScaleUncer1,LeptonScaleUncer2}, and trigger efficiencies are corrected to account for differences with respect to data. 

\subsection{Reducible background processes}

A ``reducible'' process has at least one ``fake'' lepton, that is either a lepton from a semi-leptonic decay of a $b$- or $c$-quark, referred to as heavy-flavour,  or an electron from an isolated single-track photon conversion. The contribution from misidentified light-flavour quark or gluon jets is found to be negligible based on  studies in simulation. 
The reducible background includes $WZ$, $t\bar{t}$, $t\bar{t}W$, $WW$, single $t$-quark, or single $Z$-boson production, in all cases produced in association with jets or photons. The yield of $W$ bosons with three fake leptons is negligible. 
The $t\bar{t}$ and the $WZ$ backgrounds correspond respectively to 59\% and 35\%  of the reducible background in SR1, and to 46\% each in SR2.  In both SR1 and SR2, fake leptons are predominantly fake electrons (99\%), originating either from $b$-quarks in $t\bar{t}$ candidate events or from conversions in $WZ$ candidate events. Fake muons from $b$-quark decays in $t\bar{t}$ events are suppressed by the object separation scheme described in section~\ref{sec:presel}. Regardless of the origin, the misidentification probability decreases as the lepton $p_{\rm T}$ increases.

The reducible background is estimated using a weighting method applied to events containing signal leptons ($\ell_S$) and loose leptons ($\ell_L$), which are tagged leptons failing the signal lepton requirements. 
Since the reducible background is dominated by events with at most two fake leptons, it is estimated as:
\begin{eqnarray}
&&\left[N_{\rm data}(3\ell_S+\ell_L)-N_{\rm MCirr}(3\ell_S+\ell_L)\right]\times F(\ell_L) \nonumber\\
&&-\left[N_{\rm data}(2\ell_S+\ell_{L_1}+\ell_{L_2})-N_{\rm MCirr}(2\ell_S+\ell_{L_1}+\ell_{L_2})\right]\times F(\ell_{L_1}) \times F(\ell_{L_2}),
\end{eqnarray}
where the second term corrects for the double counting of reducible-background events with two fake leptons in the first term.
The term $N_{\rm data}(3\ell_S+\ell_L)$ is the total number of events with three signal and one loose lepton, 
while $N_{\rm MCirr}(3\ell_S+\ell_L)$ is the irreducible contribution of events obtained from simulation. The definitions of $N_{\rm data}(2\ell_S+\ell_{L_1}+\ell_{L_2})$ and $N_{\rm MCirr}(2\ell_S+\ell_{L_1}+\ell_{L_2})$ are analogous.
As a conservative approach, the potential signal contamination in the $3\ell_S+\ell_L$ and $2\ell_S+\ell_{L_1}+\ell_{L_2}$ loose lepton data samples is not taken into account. 
The average ``fake ratio'' $F$ depends on the flavour and kinematics of the loose lepton $\ell_L$ and it is defined as: 

\begin{equation}
F=\sum_{i, j} \left( \alpha^{i}\times R^{ij} \times f^{ij} \right),
\end{equation}
\noindent where $i\,$ is the type of fake (heavy-flavour leptons or conversion electrons) and $j\,$ is the process category the fake originates from (top quark or $W/Z$ boson).
The fake ratios $f^{ij}$ are defined as the ratios of the probabilities that fake tagged leptons are identified as signal leptons to the probabilities that they are identified as loose leptons.
The $f^{ij}$ are determined for each relevant fake type and for each reducible-background process, and they are parameterised in muon (electron) $p_{\rm T}$ ($E_{\rm T}$) and $\eta$. 
The fake ratios are weighted according to the fractional contribution of the process they originate from through $R^{ij}$ fractions.
Both $f^{ij}$ and  $R^{ij}$ are determined in simulation.
Each correction factor $\alpha^{i}$ is the fake ratio measured in data divided by that in simulation, in control samples described below.
The fake ratios $F$ are estimated to vary from 0.8 to 0.05 for muons when the $p_{\rm T}$ increases from 10 to 100$\,$GeV. For electrons, there is little $p_{\rm T}$ dependence, and the $F$ values are $\sim$0.3.

The correction factor for heavy-flavour fakes  is measured in a $b\bar{b}$-dominated  control sample. 
This is defined by selecting events with only one $b$-tagged jet (containing a muon) and a tagged lepton. The non-$b\bar{b}$ contributions from the single and pair production of top quarks and $W$ bosons produced in association with $b$-quarks are suppressed with $\met\,$$<\,$40$\,$GeV and transverse mass $m_{\rm T}\,$$=\,$$\sqrt{2 \cdot \met \cdot p^\ell_{\rm T}  \cdot (1 - \cos\Delta\phi_{\ell, \met})}\,$$<\,$40$\,$GeV requirements, where $\Delta\phi_{\ell,\met}$ is the azimuthal angle between the tagged lepton $\ell$ and the $\met$.
The remaining non-$b\bar{b}$ background ($\sim$1\% level) is subtracted from the control sample in data using MC predictions. In this control sample, the fake ratio from heavy-flavour decays is calculated using the ratio of tagged leptons passing signal lepton requirements to those that fail. 
The correction factors are found to be $0.97\pm0.13$ and $0.86\pm0.13$ for electrons and muons respectively.

The correction factor for electron candidates originating from photon conversions is determined in a sample of photons radiated from a muon in $Z\rightarrow\mu\mu$ decays.
Events with two opposite-sign muons and one tagged electron are selected and the invariant mass of the $\mu\mu e$ triplet is required to lie within 10$\,$GeV of the nominal $Z$-boson mass value.
In this control sample, the fake ratio for conversion electrons is calculated using the ratio of tagged electrons identified as signal electrons to those identified as loose electrons.
The correction factor is found to be $1.24\pm0.13$.

\section{Systematic uncertainties}
Several sources of systematic uncertainty are considered in the signal, control and validation regions. Correlations of systematic uncertainties between processes and regions are accounted for.

MC-based sources of systematic uncertainty affect the irreducible background, the $R^{ij}$ fractions of the average fake ratios, and the signal yields.
The MC-based  systematic sources include the acceptance uncertainty due to the PDFs and the theoretical cross-section uncertainties due to the renormalisation/factorisation scale and PDFs.
The uncertainty due to the PDF set is determined using the error set of the original PDF and the scale uncertainties are calculated by varying the factorisation and renormalisation scales.
Additional systematic uncertainties are those resulting from the jet energy scale~\cite{JESuncert1} and resolution~\cite{JERuncert1}, the lepton efficiencies~\cite{LeptonScaleUncer1,LeptonScaleUncer2}, energy scales and resolutions, and uncertainties in $b$-tagging rates~\cite{btagUncer1,btagUncer2,btagUncer3,btagUncer4}.
The choice of MC generator is also included for the irreducible background.
The systematic uncertainty on the luminosity (3.9\%)~\cite{lumiconf,lumipaper} affects only the yields of the irreducible background and the signal yield.

In SR1, the total uncertainty on the irreducible background is 70\%.
This is dominated by the uncertainty on the efficiency times acceptance of the signal region selection for the $ZZ$ MC event generator, determined by comparing the \Sherpa and \POWHEG~\cite{powheg1, powheg2, powheg3, powheg4} generators and found to be 65\% of the \Sherpa $ZZ$ yield.
The next largest uncertainties on the irreducible background are due to the jet energy scale (53\%) and that due to the limited number of MC events generated (34\%).
All the remaining uncertainties in this signal region lie in the range of 0.1--5\%.
In SR2, the uncertainties are similar, except for smaller uncertainties due to the jet energy scale (3\%) and the limited number of generated events (17\%).

For the average fake ratio $F$,  the $R^{ij}$ fractions are varied between 0\% and 100\% to account for the uncertainty on the $R^{ij}$ fractions  from the sources listed above. 
The fake ratios $f^{ij}$ are generally found to be similar across processes and types of fakes in most of the kinematic regions. 
Also included in the uncertainty on the reducible background is the uncertainty from the dependence of the fake ratio on $\met$ (10\%) and the uncertainty on the fake-ratio correction factors (15--34\%).
In SR1, the dominant uncertainty on the reducible background is that due to the $R^{ij}$ fractions (75\%).
This relatively large uncertainty is due to the limited number of events in the loose lepton data sample, and the fact that those leptons fall in kinematic regions in which the fake ratio has larger variations between processes and types of fakes. 
The next to leading uncertainties are those from the statistical uncertainty on the data (60\%) and the dependence of the fake ratio on $\met$ (25\%).
In SR2, the uncertainty on the reducible background is dominated by that from the limited number of data events (140\%), followed by that due to the $R^{ij}$ fractions (6\%) and the dependence of the fake ratio on $\met$ (4\%). 

The total uncertainties on the signal yields are 10--20\% and are calculated using the method described in ref.~\cite{SUSYXS:2012}.
Signal cross-sections are calculated to NLO in the strong coupling constant using \Prospino\footnote{The addition of the resummation of soft gluon emission at NLL~\cite{Beenakker:1996ch,Kulesza:2008jb,Kulesza:2009kq,Beenakker:2009ha,Beenakker:2011fu} is performed in the case of strong SUSY pair-production.}.
An envelope of cross-section predictions is defined using the 68\% CL intervals of the \CTEQSix~\cite{Nadolsky:2008zw} (including the $\alpha_S$ uncertainty) and \MSTW~\cite{Martin:2009iq} PDF sets, together with variations of the factorisation and renormalisation scales by factors of two or one half.
The nominal cross-section value is taken to be the midpoint of the envelope and the uncertainty assigned is half the full width of the envelope, closely following the PDF4LHC recommendations~\cite{Botje:2011sn}.

\section{Background model validation \label{sec:Bvalidation}}

The background predictions have been verified in three validation regions (VR). 
A region (VR1) selects events with three signal leptons, $\met\,$$>\,$50$\,$GeV, and vetos events with $Z$-boson candidates described in section~\ref{sec:SRsel}. In VR1 the reducible background is dominated by $t\bar{t}$ production. 
The contribution from  $ZZ$ production is validated in a region (VR2) defined by events with four leptons containing a $Z$-boson candidate. 
A region (VR3) containing events with $\met\,$$<\,$50$\,$GeV, $m_{\rm eff}\,$$<\,$300$\,$GeV, four signal leptons, and no $Z$-boson candidate is used to validate the sum of the reducible and irreducible backgrounds.
The data and predictions are in agreement within the quoted statistical and systematic uncertainties, as shown in table~\ref{tab:VRs}.
Reported are the probabilities that the background fluctuates to the observed number of events or higher ($p_0$-value) and the corresponding number of standard deviations ($\sigma$). 

\begin{table}[!ht]
 \begin{center}
  \caption{Expected number of events from SUSY signals, SM backgrounds, and observed number of events in data in validation regions VR1, VR2 and VR3 (4.7$\,$fb$^{-1}$). ``SUSY ref.\ point 1'' refers to a simplified model with $m_{\chinoone} = $ 500$\,$GeV, $m_{\ninoone} = $ 300$\,$GeV, while ``SUSY ref.\ point 2'' refers to a MSUGRA model with $m_{1/2}$ = 860$\,$GeV and $\tan\beta$ = 37. 
  The uncertainties quoted include statistical and systematic effects.
  The processes labelled with ($\dagger$) are classified as irreducible background in the three-lepton validation regions.
  In the four-lepton validation regions they are included in the reducible background. The subtraction of the 
 double-counted contributions can lead to negative central values for the reducible background. \label{tab:VRs}}
 \small{
\renewcommand\arraystretch{1.2}
    \begin{tabular}{ c c c c }
    \hline    \hline
    Selection & VR1 & VR2 & VR3 \\ 
\hline
SUSY ref.\ point 1 & $4.6\pm0.5$ & $1.38\pm0.16$ & $0.004\pm0.006$ \\
SUSY ref.\ point 2 & $3.5\pm0.4$ & $2.32\pm0.34$ & $0.120\pm0.029$ \\
\hline
    $ZZ$             & $3.2\pm1.9$   & $38\pm7$ & $2.9\pm1.5$\\
    $\ttbar Z$     & $0.70\pm0.35$ & $0.64\pm0.32$ & $(3.5\pm3.6)\times 10^{-3}$\\
    $\ttbar WW$ & $0.08\pm0.06$   & $(1.3\pm1.0)\times 10^{-3}$ & $(2.2\pm2.1)\times 10^{-4}$\\
    $WZ$ ($\dagger$)         & $34.6\pm6$    & -- & --\\
    $\ttbar W$ ($\dagger$) & $2.6\pm0.8$    & -- & --\\
\hline
    $\Sigma$ Irreducible & $41\pm8$ & $38\pm7$ & $2.9\pm1.5$ \\
\hline
    Reducible & $95\pm32$ & $-0.25\pm0.96$ & $1.3\pm1.3$\\
\hline
    $\Sigma$ SM & $136\pm33$ & $38\pm7$ & $4.2\pm1.8$\\  
\hline
    Data & $152$ & $40$ & $2$  \\ 
\hline
    $p_0$-value ($\sigma$) & 0.33 (0.45) & 0.42 (0.21) & 0.80 ($-0.85$)\\
\hline        \hline
  \end{tabular}
  }
 \end{center}
\end{table}

\section{Results and interpretation}

\begin{figure}[!ht]
\centering
\subfigure[]{  \includegraphics[width=12cm]{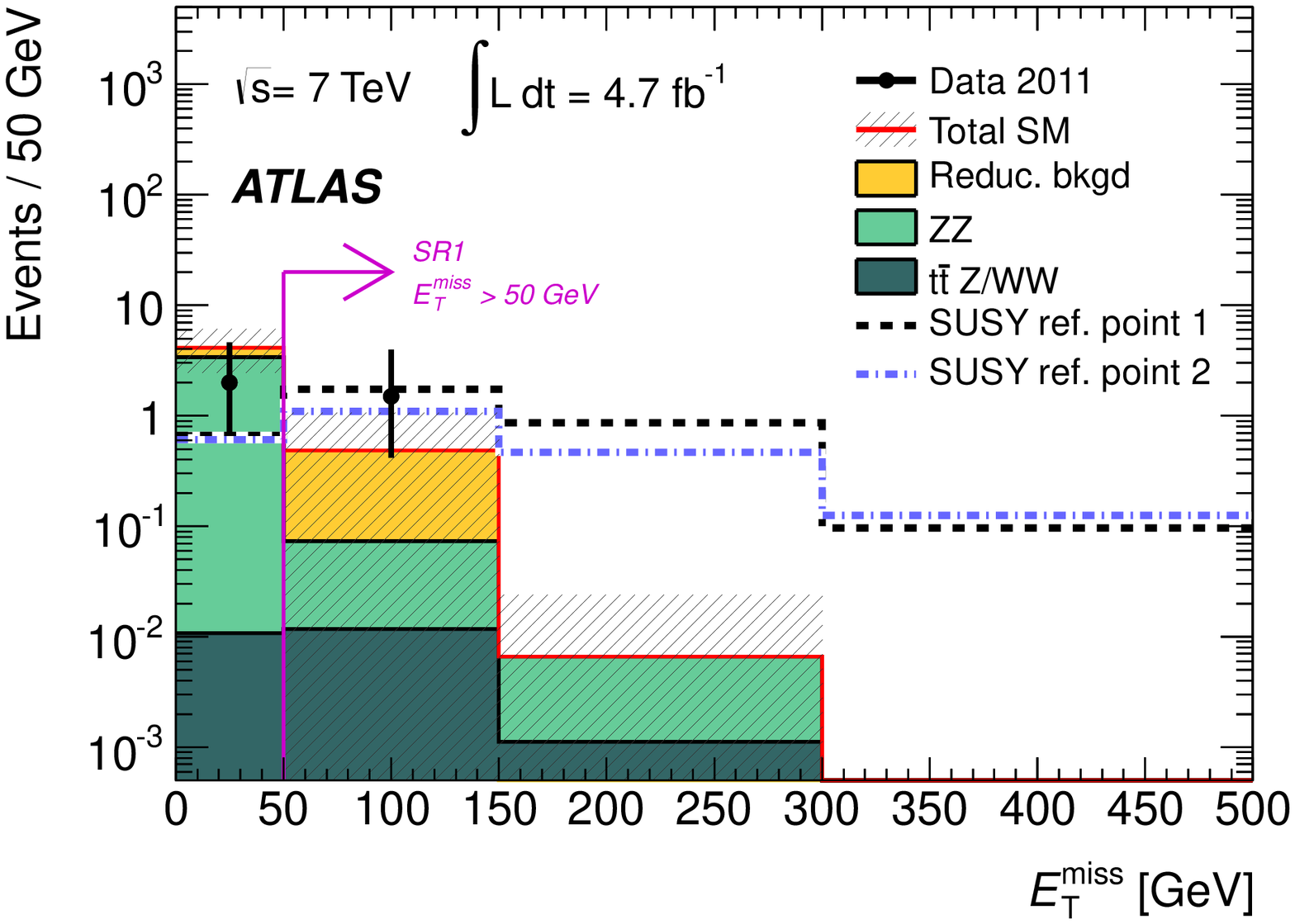}}
\subfigure[]{  \includegraphics[width=12cm]{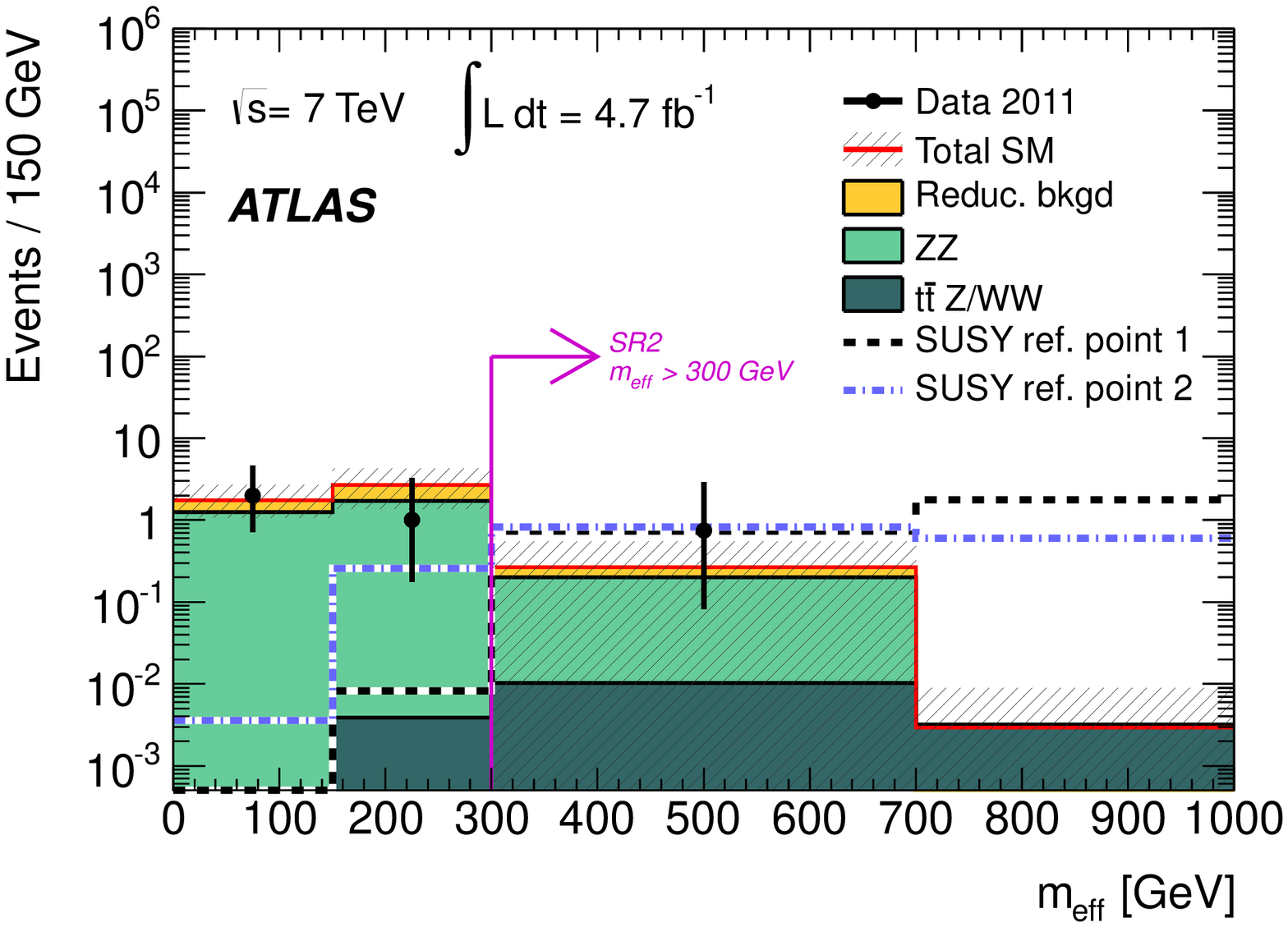}}
\caption{\label{fig:met}Distributions of (a) $\met$ and (b) $m_{\rm eff}$ for events with at least four leptons and no $Z$-boson candidates (before either the $\met$ or $m_{\rm eff}$ requirements). 
The uncertainty band includes both statistical and systematic uncertainty. 
The yields of two benchmark SUSY models are shown for illustration purposes. ``SUSY ref.\ point 1'' is a simplified model point defined by $m_{\chinoone} = $ 500$\,$GeV and $m_{\ninoone} = $ 300$\,$GeV, while ``SUSY ref.\ point 2'' is a MSUGRA model point defined by $m_{1/2}\,$$=\,$860$\,$GeV and $\tan\beta\,$$=\,$37. 
The signal distributions are not stacked on top of the expected background.
}
\end{figure}

The numbers of observed and predicted events in SR1 and SR2 are reported in table~\ref{tab:SRs}. 
Distributions of $\met$ and $m_{\rm eff}$ in events that have at least four leptons and no $Z$-boson candidates (before either the $\met$ or $m_{\rm eff}$ requirements) are presented in fig.~\ref{fig:met}. 

\begin{table}[!ht]
 \begin{center}
  \caption{Expected number of events from SUSY signals, SM backgrounds, and observed number of events in data in signal regions SR1 and SR2 (4.7$\,$fb$^{-1}$). 
``SUSY ref.\ point 1'' refers to a simplified model with $m_{\chinoone} = $ 500$\,$GeV, $m_{\ninoone} = $ 300$\,$GeV, while ``SUSY ref.\ point 2'' refers to a MSUGRA model with $m_{1/2}$ = 860$\,$GeV and $\tan\beta$ = 37. 
 The uncertainties quoted include statistical and systematic effects.
 Upper limits on the observed (expected) visible production cross-section of new physics processes at 95\% CL are also shown. \label{tab:SRs}
  }
  \small{
\renewcommand\arraystretch{1.2}
    \begin{tabular}{ c c c }
    \hline    \hline
    Selection & SR1 & SR2 \\
\hline
SUSY ref.\ point 1 & $6.5\pm0.6$ & $7.1\pm0.7$ \\
SUSY ref.\ point 2 & $4.2\pm0.6$ & $4.5\pm0.6$ \\
\hline
    $ZZ$ & $0.14\pm0.11$ & $0.51\pm0.30$ \\
    $\ttbar Z$ & $0.023\pm0.014$ & $0.029\pm0.016$\\
    $\ttbar WW$ & $0.0044\pm0.0035$ & $0.005\pm0.004$\\
\hline
    $\Sigma$ Irreducible & $0.17\pm0.12$ & $0.54\pm0.31$ \\
\hline
    Reducible & $0.8\pm0.8$ & $0.18\pm0.26$\\
\hline
  {\bf  $\Sigma$ SM} &${\bf 1.0\pm0.8}$ &${\bf  0.7\pm0.4}$\\  
\hline
  {\bf  Data }& ${\bf 3 }$&${\bf 2 }$ \\ 
\hline
    $p_0$-value ($\sigma$) & 0.05 (1.7) & 0.07 (1.5)\\
\hline
    $\sigma_{\rm vis}$ obs (exp) & 1.3 (0.8) & 1.1 (0.7)\\
\hline        \hline
  \end{tabular}
  }
 \end{center}
\end{table}

No significant excess of events is found in the signal regions. 
Upper limits on the visible cross-section of new physics processes are calculated, defined by the product of production cross-section, acceptance and efficiency, and placed at 95\% CL with the modified frequentist CL$_s$ prescription~\cite{CLs}. All systematic uncertainties and their correlations are taken into account via nuisance parameters in a profile likelihood fit~\cite{cowan}. Observed 95\% CL limits on the visible cross-section of new physics processes are placed at 1.3$\,$fb in SR1 and 1.1$\,$fb in SR2. The corresponding expected limits are 0.8$\,$fb and 0.7$\,$fb, respectively. 

\begin{figure}[!ht]
\centering
\subfigure[]{  \includegraphics[width=9cm]{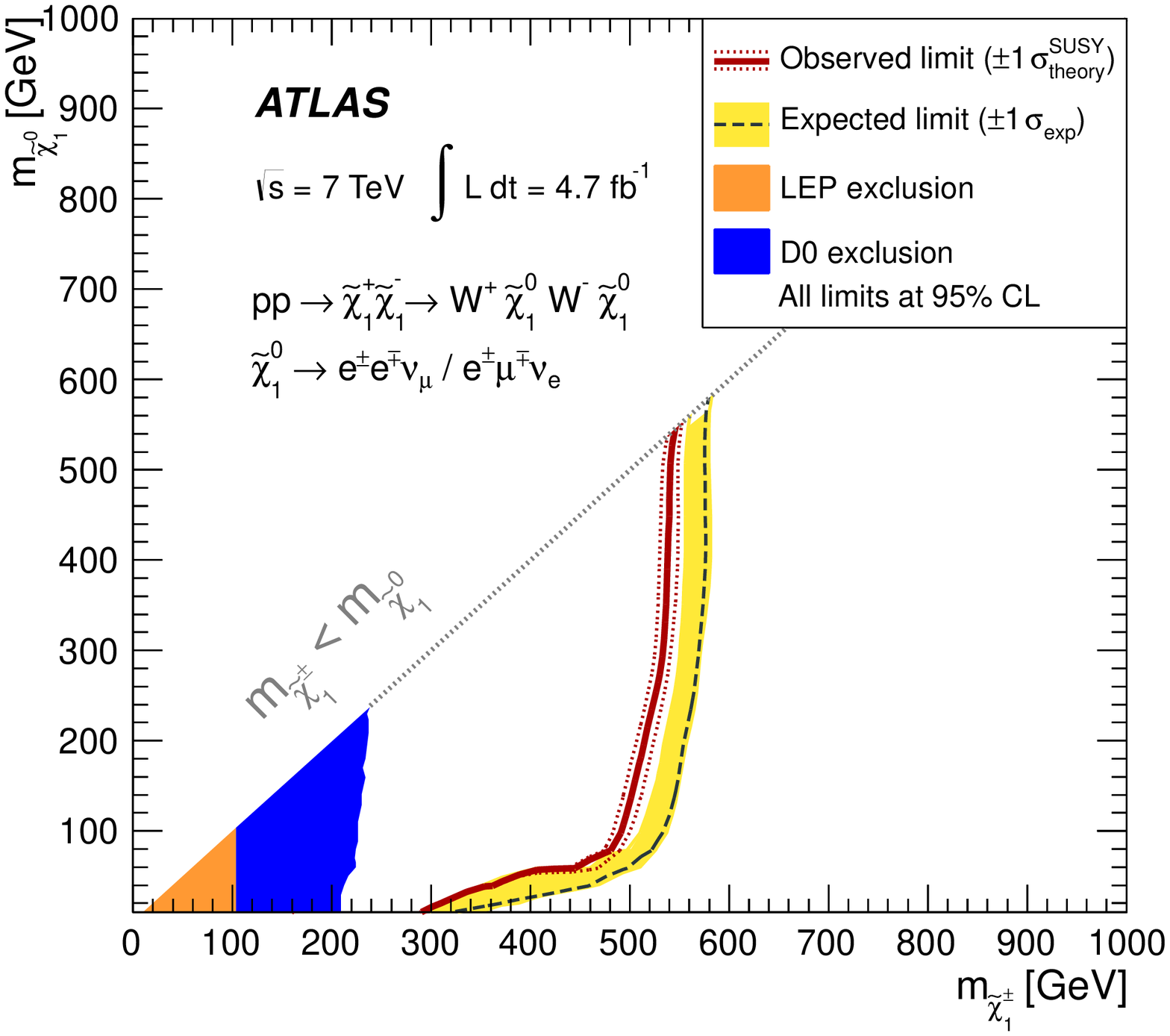}}
\subfigure[]{  \includegraphics[width=9cm]{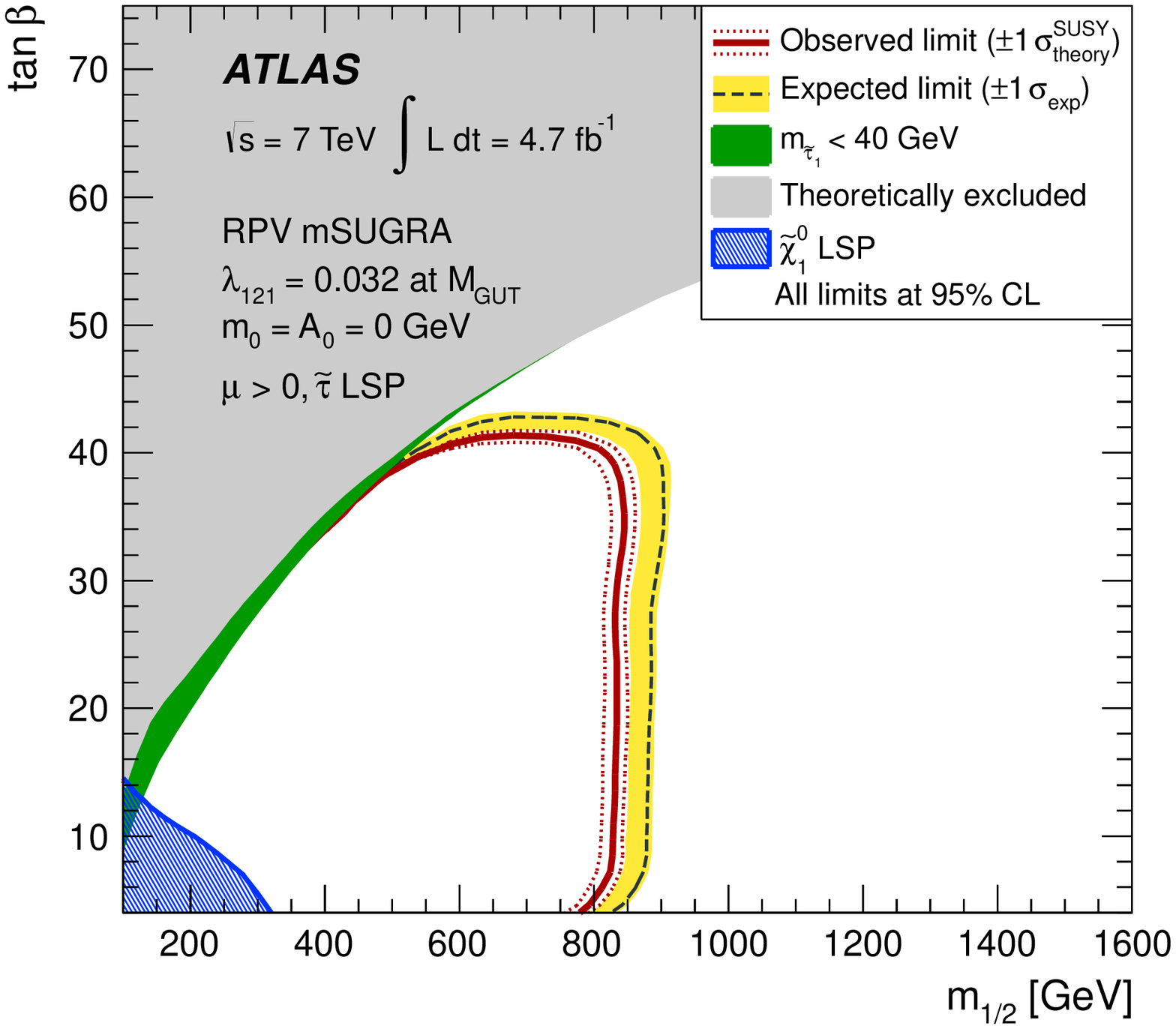}}
\caption{Observed and expected 95\% CL limit contours for (a) simplified model and (b) MSUGRA/CMSSM. 
The expected and observed limits are calculated without signal cross-section uncertainty taken into account.
The yellow band is the $\pm$1$\sigma$ experimental uncertainty on the expected limit (black dashed line).
The red dotted lines are the $\pm$1$\sigma$ signal theory uncertainty on the observed limit (red solid line). 
Linear interpolation is used to account for the discreteness of the signal grids.
The exclusion contours are optimised by applying in each signal grid point the CL values from the more sensitive signal region (lowest expected CL). 
It should be noted that the $y$-axis does not start at zero for either plot. 
The LEP limit corresponds to the limit on the $\chinoone$ mass in ref.~\cite{PDG}, while the D0 limit corresponds to the limit in the $\chinoone,\ninoone$ mass plane in ref.~\cite{D02006}. The hard $m_{\stau}\,$$<\,$40$\,$GeV limit is from the LEP $Z$ width measurement~\cite{PDGstaumass}.
\label{fig:limits}
}
\end{figure}

The results of the analysis are interpreted in two RPV SUSY scenarios and shown in fig.~\ref{fig:limits}, where the limits are calculated choosing the signal region with the best expected limit for each of the model points. 
The uncertainties on the signal cross-section are not included in the limit calculation but their impact on the observed limit is shown by the red dotted lines.

The main features of the exclusion limits can be explained in broad terms as follows.
In the simplified model the sensitivity is governed by the production cross-section, which decreases at large $m_{\chinoone}$ values, and by the efficiency of the signal region selection cuts. 
The requirements on the minimum value of the SFOS dilepton mass, and on the minimum lepton-lepton separation reduce the acceptance and selection efficiency for leptons from light $\ninoone$, hence only values of 10$\,$GeV and above are considered. As the mass of the $\ninoone$ approaches zero, the phase space for leptonic decays is greatly reduced, while values of $m_{\ninoone}\,$=$\,$0$\,$GeV are not possible without the LSP becoming stable. In the MSUGRA/CMSSM model, the production cross-section  decreases as $m_{1/2}$ increases, leading to  smaller sensitivities in the high-mass region. 
Due to lower values of the four-body decay branching ratio and an increased $\stau$ lifetime, the sensitivity drops for $\tan\beta$ values above 40. 
In the region of the parameter space with $m_{1/2}\sim\,$800 GeV, weak gaugino production contributes 50\% to the total SUSY production cross-section, while $\tilde{e}/\tilde{\mu}/\tilde{\nu}$ ($\stau$) production contributes 10--20\% (10--30\%), and strong production 20\%. Similar fractional contributions are also seen in the two signal regions.

\section{Summary}

Results from a search for new phenomena in the final state with four or more leptons (electrons or muons) and either moderate values of missing transverse momentum or large effective mass are reported.
The analysis is based on 4.7$\,$fb$^{-1}$ of proton-proton collision data delivered by the LHC at $\sqrt{s}\,$$=\,$7$\,$TeV. 
No significant excess of events is found in data. Observed 95\% CL limits on the visible cross-section  are placed at 1.3$\,$fb and 1.1$\,$fb in the two signal regions, respectively. 
The null result is interpreted in a simplified model of chargino pair-production in which each chargino cascades to the lightest neutralino that decays into two charged leptons ($ee$ or $e\mu$) and a neutrino via an RPV coupling. 
In the simplified model of RPV supersymmetry, chargino masses up to 540$\,$GeV are excluded for LSP masses above 300$\,$GeV.
Limits are also set in an RPV MSUGRA model with a $\tilde{\tau_{1}}$ LSP that promptly decays into a $\tau$ lepton, two charged leptons and a neutrino, where values of $m_{1/2}$ below 820$\,$GeV are excluded when $10<\tan\beta<40$.

\FloatBarrier

\section{Acknowledgements}
We thank CERN for the very successful operation of the LHC, as well as the
support staff from our institutions without whom ATLAS could not be
operated efficiently.

We acknowledge the support of ANPCyT, Argentina; YerPhI, Armenia; ARC,
Australia; BMWF and FWF, Austria; ANAS, Azerbaijan; SSTC, Belarus; CNPq and FAPESP,
Brazil; NSERC, NRC and CFI, Canada; CERN; CONICYT, Chile; CAS, MOST and NSFC,
China; COLCIENCIAS, Colombia; MSMT CR, MPO CR and VSC CR, Czech Republic;
DNRF, DNSRC and Lundbeck Foundation, Denmark; EPLANET and ERC, European Union;
IN2P3-CNRS, CEA-DSM/IRFU, France; GNSF, Georgia; BMBF, DFG, HGF, MPG and AvH
Foundation, Germany; GSRT, Greece; ISF, MINERVA, GIF, DIP and Benoziyo Center,
Israel; INFN, Italy; MEXT and JSPS, Japan; CNRST, Morocco; FOM and NWO,
Netherlands; BRF and RCN, Norway; MNiSW, Poland; GRICES and FCT, Portugal; MERYS
(MECTS), Romania; MES of Russia and ROSATOM, Russian Federation; JINR; MSTD,
Serbia; MSSR, Slovakia; ARRS and MVZT, Slovenia; DST/NRF, South Africa;
MICINN, Spain; SRC and Wallenberg Foundation, Sweden; SER, SNSF and Cantons of
Bern and Geneva, Switzerland; NSC, Taiwan; TAEK, Turkey; STFC, the Royal
Society and Leverhulme Trust, United Kingdom; DOE and NSF, United States of
America.

The crucial computing support from all WLCG partners is acknowledged
gratefully, in particular from CERN and the ATLAS Tier-1 facilities at
TRIUMF (Canada), NDGF (Denmark, Norway, Sweden), CC-IN2P3 (France),
KIT/GridKA (Germany), INFN-CNAF (Italy), NL-T1 (Netherlands), PIC (Spain),
ASGC (Taiwan), RAL (UK) and BNL (USA) and in the Tier-2 facilities
worldwide.

\bibliographystyle{JHEP}
\raggedright
\bibliography{paper}

\providecommand{\href}[2]{#2}\begingroup\raggedright\begin{thebibliography}{10}

\bibitem{Miyazawa:1966}
{H. Miyazawa,} {\em Prog. Theor. Phys.} {\bf 36} (1966) 1266.

\bibitem{Ramond:1971gb}
{P. Ramond,} {\em Phys. Rev.} {\bf D 3} (1971) 2415.

\bibitem{Golfand:1971iw}
{Y. Golfand, E. Likhtman,} {\em JETP Lett.} {\bf 13} (1971) 323.

\bibitem{Neveu:1971rx}
{A. Neveu, J. H. Schwarz,} {\em {Nucl. Phys.}} {\bf {B 31}} ({1971}) {86}.

\bibitem{Neveu:1971iv}
{A. Neveu, J. H. Schwarz,} {\em {Phys. Rev.}} {\bf {D 4}} ({1971}) {1109}.

\bibitem{Gervais:1971ji}
{J. L. Gervais, B. Sakita,} {\em {Nucl. Phys.}} {\bf {B 34}} ({1971}) 632.

\bibitem{Volkov:1973ix}
{D. Volkov, V. Akulov,} {\em Phys. Lett.} {\bf B 46} (1973) 109.

\bibitem{Wess:1973kz}
{J. Wess, B. Zumino,} {\em Phys. Lett.} {\bf B 49} (1974) 52.

\bibitem{Wess:1974tw}
{J. Wess, B. Zumino,} {\em Nucl. Phys.} {\bf B 70} (1974) 39.

\bibitem{technicolor}
{E.~Eichten, K.~Lane,} {\em Phys. Lett.} {\bf B 90} (1980) 125.

\bibitem{extraBoson}
{P.~Langacker,} {\em Rev.~Mod.~Phys} {\bf 81} (2009) 1199, arXiv:0801.1345.

\bibitem{heavyNeutrino}
{F.~del~Aguila, J.~Aguilar-Saavedra,} {\em Nucl. Phys.} {\bf B 813} (2009) 22,
  arXiv:0808.2468.

\bibitem{D02006}
{D0 Collaboration, V.~M.~Abazov {\it et al.},} {\em Phys. Lett.} {\bf B 638}
  (2006) 441, arXiv:hep--ex/0605005.

\bibitem{CDF2007}
{CDF Collaboration, A.~Abulencia {\it et al.},} {\em Phys. Rev. Lett.} {\bf 98}
  (2007) 131804, arXiv:0706.4448.

\bibitem{CMS4L}
{CMS Collaboration,} {\em JHEP} {\bf 06} (2012) 169, arXiv:1204.5341.

\bibitem{protonDecay}
{Super-Kamiokande Collaboration, C.~Regis {\it et al.},} {\em Phys.~Rev.} {\bf
  D 86} (2012) 012006.

\bibitem{Fayet:1976et}
{P. Fayet,} {\em Phys. Lett.} {\bf B 64} (1976) 159.

\bibitem{Fayet:1977yc}
{P. Fayet,} {\em Phys. Lett.} {\bf B 69} (1977) 489.

\bibitem{Farrar:1978xj}
{G.~R.~Farrar, P.~Fayet,} {\em Phys. Lett.} {\bf B 76} (1978) 575.

\bibitem{Fayet:1979sa}
{P. Fayet,} {\em Phys. Lett.} {\bf B 84} (1979) 416.

\bibitem{Dimopoulos:1981zb}
{S.~Dimopoulos, H.~Georgi,} {\em Nucl. Phys.} {\bf B 193} (1981) 150.

\bibitem{herbiMSSMsymm}
{H.~K.~Dreiner, X.~L.~Luhn, M.~Thormeier,} {\em Phys. Rev.} {\bf D 73} (2006)
  075007, arXiv:hep--ph/0512163.

\bibitem{neutrinomass_benall}
{B.~C.~Allanach, C.~H.~Kom}, {arXiv:0712.0852, 2007}.

\bibitem{simplModels1}
{J. Alwall, P.~Schuster, N.~Toro,} {\em {Phys. Rev.}} {\bf {D 79}} ({2009})
  {075020, arXiv:0810.3921}.

\bibitem{Desch:2011}
{ K.~Desch {\it et al.},} {\em Phys. Rev} {\bf D 83} (2011) 015013.

\bibitem{bintosgamma}
{B.~Allanach {\it et al.},} {\em Phys. Rev.} {\bf D 75} (2007) 035002,
  arXiv:hep--ph/0609263.

\bibitem{Kao:2009fg}
{Y.~Kao, T.~Takeuchi}, {arXiv:0910.4980, 2009}.

\bibitem{HiggsATLAS}
{ATLAS Collaboration}, {accepted by {\it Phys. Lett.} {\bf B}, arXiv:1207.7214,
  2012}.

\bibitem{HiggsCMS}
{CMS Collaboration}, {accepted by {\it Phys. Lett.} {\bf B}, arXiv:1207.7235,
  2012}.

\bibitem{atlas-det}
{ATLAS Collaboration,} {\em JINST} {\bf 3} (2008) S08003.

\bibitem{SherpaMC}
{T.~Gleisberg {\it et al.},} {\em JHEP} {\bf 0902} (2009) 007, arXiv:0811.4622.

\bibitem{madgraph}
{J. Alwall {\it et al.},} {\em JHEP} {\bf 06} (2011) 128, arXiv:1106.0522.

\bibitem{mcatnlo}
{S. Frixione, B.~R.~Webber,} {\em JHEP} {\bf 0206} (2002) 029,
  arXiv:hep--ph/0204244.

\bibitem{alpgen}
{M. Mangano {\it et al.},} {\em JHEP} {\bf 0307} (2003) 001,
  arXiv:hep--ph/0206293.

\bibitem{mcfm1}
{J. M. Campbell, R. K. Ellis,} {\em Phys. Rev.} {\bf D 60} (1999) 113006,
  arXiv:hep--ph/9905386.

\bibitem{mcfm2}
{J. M. Campbell, R. K. Ellis, C. Williams,} {\em JHEP} {\bf 07} (2011) 018,
  arXiv:1105.0020.

\bibitem{hathor}
{M.~Aliev {\it et al.},} {\em {Comput. Phys. Commun.}} {\bf {182}} ({2011})
  {1034, arXiv:1007.1327}.

\bibitem{ttZ}
{A. Kardos {\it et al.}}, {arXiv:1111.0610, 2011}.

\bibitem{ttW}
{J. M. Campbell, R. K. Ellis}, {arXiv:1204.5678, 2012}.

\bibitem{wgamma}
{ATLAS Collaboration}, {submitted to {\it Phys. Lett.} {\bf B},
  arXiv:1205.2531, 2012}.

\bibitem{fewz1}
{K. Melnikov, F. Petriello,} {\em Phys. Rev.} {\bf D 74} (2006) 114017,
  arXiv:hep--ph/0609070.

\bibitem{fewz2}
{C. Anastasiou {\it et al.},} {\em {Phys. Rev.}} {\bf {D 69}} ({2004})
  {094008}.

\bibitem{CTEQ6L1}
{P.~M.~Nadolsky {\it et al.},} {\em Phys. Rev.} {\bf D 78} (2008) 013004,
  arXiv:0802.0007.

\bibitem{CT10pdf}
{H. Lai {\it et al.},} {\em Phys. Rev.} {\bf D 82} (2010) 074024,
  arXiv:1007.2241.

\bibitem{herwigplusplus}
{M.~Bahr {\it et al.}} {\em {Eur. Phys. J.}} {\bf C 58} (2008) 639,
  arXiv:0803.0883.

\bibitem{herwig}
{G.~Corcella {\it et al.},} {\em JHEP} {\bf 0101} (2001) 010,
  arXiv:hep--ph/0011363.

\bibitem{Beenakker:1996ch}
{W.~Beenakker {\it et al.},} {\em Nucl. Phys.} {\bf B492} (1997) 51,
  arXiv:hep--ph/9610490.

\bibitem{pythia}
{T.~Sjostrand, S.~Mrenna, P.~Skands,} {\em JHEP} {\bf 05} (2006) 026,
  arXiv:hep--ph/0603175.

\bibitem{jimmy}
{J. M. Butterworth, J. R. Forshaw, M. H. Seymour,} {\em Z. Phys.} {\bf C 72}
  (1996) 637, arXiv:1005.4568.

\bibitem{Agostinelli:2002hh}
{GEANT4 Collaboration, S. Agostinelli {\it et al.},} {\em Nucl. Instrum. Meth.}
  {\bf A 506} (2003) 250.

\bibitem{simulation}
{ATLAS Collaboration,} {\em {Eur. Phys. J}} {\bf {C 70}} ({2010}) 823,
  arXiv:1005.4568.

\bibitem{electronPerf}
{ATLAS Collaboration,} {\em Eur. Phys. J} {\bf C 72} (2012) 1909,
  arXiv.:1110.3174.

\bibitem{muon}
{ATLAS Collaboration,} {\em JHEP} {\bf 12} (2010) 60.

\bibitem{Cacciari:2008gp}
{M.~Cacciari, G.~P.~Salam, G.~Soyez,} {\em JHEP} {\bf 04} (2008) 063,
  arXiv:0802.1189.

\bibitem{fakeCleanUp}
{ATLAS Collaboration}, submitted to {\it Eur. Phys. J.} {\bf C},
  arXiv:1112.6426, 2011.

\bibitem{btag}
{ATLAS Collaboration}, {ATLAS-CONF-2011-102,
  http://cdsweb.cern.ch/record/1369219, 2011}.

\bibitem{metPerf}
{ATLAS Collaboration,} {\em Eur. Phys. J.} {\bf {C 72}} ({2012}) {1844,
  arXiv:1108.5602}.

\bibitem{JESuncert1}
{ATLAS Collaboration}, {submitted to {\it Eur. Phys. J.} {\bf C},
  arXiv:1112.6426, 2011}.

\bibitem{JERuncert1}
{ATLAS Collaboration}, {ATLAS-CONF-2010-054,
  http://cdsweb.cern.ch/record/1337782, 2010}.

\bibitem{LeptonScaleUncer1}
{ATLAS Collaboration}, {ATLAS-CONF-2011-063,
  http://cdsweb.cern.ch/record/1345743, 2011}.

\bibitem{LeptonScaleUncer2}
{ATLAS Collaboration}, {ATLAS-CONF-2011-021,
  http://cdsweb.cern.ch/record/1336750, 2011}.

\bibitem{btagUncer1}
{ATLAS Collaboration}, {ATLAS-CONF-2011-102,
  http://cdsweb.cern.ch/record/1369219, 2011}.

\bibitem{btagUncer2}
{ATLAS Collaboration}, {ATLAS-CONF-2011-143,
  http://cdsweb.cern.ch/record/1376703, 2011}.

\bibitem{btagUncer3}
{ATLAS Collaboration}, {ATLAS-CONF-2012-039,
  http://cdsweb.cern.ch/record/1435193, 2012}.

\bibitem{btagUncer4}
{ATLAS Collaboration}, {ATLAS-CONF-2012-040,
  http://cdsweb.cern.ch/record/1435194, 2012}.

\bibitem{lumiconf}
{ATLAS Collaboration}, {ATLAS-CONF-2011-116,
  http://cdsweb.cern.ch/record/1376384, 2011}.

\bibitem{lumipaper}
{ATLAS Collaboration,} {\em {Eur. Phys. J. }} {\bf {C 71}} (2011) {1630,
  arXiv:1101.2185}.

\bibitem{powheg1}
{T.~Melia {\it et al.},} {\em JHEP} {\bf 1111} (2011) 078, arXiv:1107.5051.

\bibitem{powheg2}
{P.~Nason,} {\em JHEP} {\bf 0411} (2004) 040, arXiv:hep--ph/0409146.

\bibitem{powheg3}
{S.~Frixione {\it et al.},} {\em JHEP} {\bf 0711} (2007) 070, arXiv:0709.2092.

\bibitem{powheg4}
{S.~Alioli {\it et al.},} {\em JHEP} {\bf 1006} (2010) 043, arXiv:1002.2581.

\bibitem{SUSYXS:2012}
{M.~Kramer {\it et al.}}, {CERN-PH-TH/2012-163, arXiv:1206.2892, 2012}.

\bibitem{Kulesza:2008jb}
{A. Kulesza and L. Motyka,} {\em Phys. Rev. Lett.} {\bf 102} (2009) 111802,
  arXiv:0807.2405.

\bibitem{Kulesza:2009kq}
{A. Kulesza and L. Motyka,} {\em Phys. Rev.} {\bf D 80} (2009) 095004,
  arXiv:0905.4749.

\bibitem{Beenakker:2009ha}
{W.~Beenakker {\it et al.},} {\em JHEP} {\bf 0912} (2009) 041, arXiv:0909.4418.

\bibitem{Beenakker:2011fu}
{W.~Beenakker {\it et al.},} {\em Int. J. Mod. Phys.} {\bf A 26} (2011) 2637,
  arXiv:1105.1110.

\bibitem{Nadolsky:2008zw}
{P. M. Nadolsky {\it et al.},} {\em Phys. Rev.} {\bf D 78} (2008) 013004.

\bibitem{Martin:2009iq}
{A. D. Martin {\it et al.},} {\em Eur. Phys. J.} {\bf C 63} (2009) 189.

\bibitem{Botje:2011sn}
{M. Botje {\it et al.}}, arXiv:1101.0538, 2011.

\bibitem{CLs}
{A. L. Read,} {\em J. Phys.} {\bf G 28} (2002) 2693.

\bibitem{cowan}
{G. Cowan {\it et al.},} {\em Eur. Phys. J.} {\bf C 71} (2011) 1554,
  arXiv:1007.1727.

\bibitem{PDG}
{Particle Data Group, K.~Nakamura {\it et al.},} {\em J. Phys.} {\bf G 37}
  (2010) 075021.

\bibitem{PDGstaumass}
{Particle Data Group, J.~Beringer {\it et al.},} {\em Phys. Rev.} {\bf D 86}
  (2012) 010001.

\end{thebibliography}\endgroup

\clearpage

\begin{flushleft}
{\Large The ATLAS Collaboration}

\bigskip

G.~Aad$^{\rm 48}$,
T.~Abajyan$^{\rm 21}$,
B.~Abbott$^{\rm 111}$,
J.~Abdallah$^{\rm 12}$,
S.~Abdel~Khalek$^{\rm 115}$,
A.A.~Abdelalim$^{\rm 49}$,
O.~Abdinov$^{\rm 11}$,
R.~Aben$^{\rm 105}$,
B.~Abi$^{\rm 112}$,
M.~Abolins$^{\rm 88}$
% ...
\bigskip
\\
% ...
$^{11}$ Institute of Physics, Azerbaijan Academy of Sciences, Baku, Azerbaijan\\
$^{12}$ Institut de F{\'\i}sica d'Altes Energies and Departament de F{\'\i}sica de la Universitat Aut{\`o}noma de Barcelona and ICREA, Barcelona, Spain\\
% ...
$^{21}$ Physikalisches Institut, University of Bonn, Bonn, Germany\\
% ...
$^{48}$ Fakult{\"a}t f{\"u}r Mathematik und Physik, Albert-Ludwigs-Universit{\"a}t, Freiburg, Germany\\
$^{49}$ Section de Physique, Universit{\'e} de Gen{\`e}ve, Geneva, Switzerland\\
% ...
$^{88}$ Department of Physics and Astronomy, Michigan State University, East Lansing MI, United States of America\\
% ...
$^{105}$ Nikhef National Institute for Subatomic Physics and University of Amsterdam, Amsterdam, Netherlands\\
% ...
$^{111}$ Homer L. Dodge Department of Physics and Astronomy, University of Oklahoma, Norman OK, United States of America\\
$^{112}$ Department of Physics, Oklahoma State University, Stillwater OK, United States of America\\
% ...
$^{115}$ LAL, Universit{\'e} Paris-Sud and CNRS/IN2P3, Orsay, France\\
% ...
\end{flushleft}


\end{document}